\documentclass[a4paper,11pt]{article}
\pdfoutput=1 

\usepackage{jheppub} 

\usepackage[utf8]{inputenc} %
\usepackage{xfrac}
\usepackage{ulem}   
\usepackage{xcolor}
\usepackage[toc,page]{appendix}

\newcommand{\OP}{\mathcal{O}}       
\newcommand{\crit}{\mathrm{crit}}   
\newcommand{\equ}{\!=\!}
\newcommand{\mi}{\!-\!}
\newcommand{\plu}{\!+\!}

\title{\boldmath Topology induced first-order phase transitions in lattice quantum gravity}

\author[a,b]{J. Ambjorn}
\author[c,d]{J. Gizbert-Studnicki}
\author[c,d]{A. Görlich}
\author[c]{D. Németh}

\affiliation[a]{The Niels Bohr Institute, Copenhagen University, Blegdamsvej 17, 2100 Copenhagen, Denmark}
\affiliation[b]{IMAAP, Radboud University, Nijmegen, PO Box 9010, The Netherlands}
\affiliation[c]{Institute of Theoretical Physics, Jagiellonian University,
Łojasiewicza 11, Kraków, 30-348, Poland}
\affiliation[d]{Mark Kac Complex Systems Research Center, Jagiellonian University,
Łojasiewicza 11, Kraków, 30-348, Poland}

\emailAdd{ambjorn@nbi.dk}
\emailAdd{jakub.gizbert-studnicki@uj.edu.pl}
\emailAdd{andrzej.goerlich@uj.edu.pl}
\emailAdd{nemeth.daniel.1992@gmail.com}

\abstract{ 
Causal Dynamical Triangulations (CDT) is a lattice formulation of quantum gravity, suitable for Monte-Carlo simulations which have been used to study the phase diagram of the model. It has four phases characterized by different dominant geometries, denoted phase $A$, $B$, $C$ and $C_b$. In this article we analyse the $A-B$ and the $B-C$ {phase} transitions in the case where the topology of space is that of the three-torus. This completes the phase diagram of CDT for such a spatial topology. We observe that the order of a phase transition of spacetime geometries can depend on the topology of spacetime.
}

\begin{document}

\maketitle

\section{Introduction}

\subsection{Causal Dynamical Triangulations}

Causal Dynamical Triangulations (CDT) is a lattice formulation of quantum gravity \cite{review1,review2} based on the path integral approach of Feynman, where certain piecewise linear geometries provide the lattice structure and the UV cut-off. The action used is the Einstein-Hilbert action provided for such geometries by Regge \cite{regge}. It is the hope that it can be as successful as Lattice QCD has been in regard to testing non-perturbative aspects of the theory via numerical simulations. The difference between CDT and Lattice QCD is that in the latter case the lattice structure is fixed and is just representing the underlying flat spacetime, while in CDT the geometry is encoded in the lattice connectivity. Since we want to integrate over geometries in the path integral, we have to use lattices with different connectivity structure. More precisely, within the framework of CDT, the formal path integral of quantum gravity is defined as a sum over geometries constructed by gluing together simplicial building blocks satisfying certain topological constraints. The most important constraint is that the quantum geometries admit a global time-foliation into spatial slices of fixed topology which is conserved in the time evolution. In all cases studied before, the topology of the spatial slices was chosen to be either spherical ($S^3$) or toroidal ($T^3$). Even though the formulation seems to be discrete in nature, it is important to stress that the lengths of links (i.e., the edges of the simplicial building blocks) play a role of the UV cut-off, which is supposed to be removed in the continuum limit, if it exists. In 3+1 dimensional spacetime the simplicial manifolds are constructed from two types of building blocks: the $\{4,1\}$ and the $\{3,2\}$ simplices, where the first number denotes the number of vertices at the (discrete) lattice time $t$  and the second one is the number of vertices at time $t\pm 1$. Within each four-simplex, all vertices with the same time coordinate are connected by space-like links and vertices at the neighboring time layers are connected by time-like links. The lengths of all spatial links are fixed to be $a_s$ and of all time-like links to be $a_t= - \alpha \, a_s^2$, where the asymmetry parameter $\alpha >0 $ if the signature of spacetime is Lorentzian. The simplices are then glued together along their three-dimensional faces (tetrahedra) in such a way that no topological defects are introduced and the assumed spatial topology and time-foliation structure of the simplicial manifolds are preserved. The imposed global time-foliation allows for a well-defined analytic continuation between Lorentzian and Euclidean signatures. The quantum amplitude can be calculated as a weighted sum over all such discretized four-dimensional geometries, later called \textit{triangulations} or \textit{configurations}, joining two geometric states. In practice, one usually identifies the initial and final states and thus assumes time-periodic boundary conditions, leading to the global topology of the CDT manifolds being either $T^1\times S^3$ or $T^4$, in what we call the \textit{spherical} or the \textit{toroidal} CDT, respectively, depending on the (fixed) spatial topology choice. In the continuum formulation, the path integral is formally defined  as:
\begin{equation}
    \mathcal{Z}_{\mathrm{QG}} = \int\mathcal{D}_\mathcal{M}[g]e^{iS_{EH}[g]} 
\end{equation}
where $[g]$ denotes the equivalence class of metrics with respect to diffeomorphisms and $\mathcal{D_\mathcal{M}}$[g] is the integration measure over nonequivalent classes of metrics. This expression becomes meaningful only if one introduces a regularization. In CDT we make a particular choice by introducing the piecewise linear regularization described above, and when performing the Wick rotation from Lorentzian to Euclidean signature, the path integral becomes a partition function:
\begin{equation}
\mathcal{Z}_{\mathrm{CDT}} = \sum_\mathcal{T} e^{-S_R[\mathcal{T]}}
\label{PI}
\end{equation}
with the Regge action  \cite{nonperturbative}: 
\begin{equation}
    S_R = -(\kappa_0 +6 \Delta) \cdot N_0 + \kappa_4 \cdot (N_{41} + N_{32}) + \Delta \cdot N_{41},
    \label{Action}
\end{equation}
where $N_0$ is the number of vertices in a triangulation $\mathcal{T}$, $N_{41}$ and $N_{32}$ are the numbers of $\{4,1\}$ and $\{3,2\}$ simplices, respectively. The action is parametrized by a set of three dimensionless bare coupling constants, $\kappa_0, \kappa_4$ and $\Delta$ corresponding respectively to the inverse gravitational bare coupling constant, the dimensionless cosmological constant and a function of the asymmetry parameter $\alpha$. The model is analytically solvable  in 1+1 dimensions \cite{cdt1d} and the 2+1 dimensional model can be associated with matrix models \cite{abab1,abab2,abab3}. The 3+1 dimensional model discussed here can be analyzed  via  Monte Carlo simulations.

\subsection{Phase transition methodology}\label{sec:method}

The phase structure is such that for given $\kappa_0$ and $\Delta$ there exists a $\kappa_4^{\crit}$ such that for $\kappa_4 > \kappa_4^{\crit}$ the average value of $N_4 \equ N_{41} \plu N_{32}$ will be finite, while for $\kappa_4 < \kappa_4^{\crit}$ the partition function is not well defined. We are interested in the limit where $N_4 \to \infty$. In order to obtain that in a controlled way, in the numerical simulations where $N_4$ will fluctuate, we add to the action (\ref{Action}) a quadratic volume fixing term: $\epsilon \, (N_{41}\mi \bar N_{41})^2$. Then, to achieve that $N_{41}$ actually fluctuates around $\bar{N}_{41}$ in the large-volume limit, one has to add the volume fixing term and fine-tune the bare cosmological constant $\kappa_4$ to its pseudo-critical value $\kappa_4^{\crit}(\bar{N}_{41})$. Only for $\bar{N}_{41}\to\infty$ will $\kappa_4^{\crit}(\bar{N}_{41})$ approach the true critical value $\kappa_4^{\crit}$. The nontrivial behavior of $\kappa_4$ near $\kappa_4^{\crit}$ is transferred to the volume dependence of $\kappa_4$ on $\bar{N}_{41}$. It means that for each $\bar{N}_{41}$ we effectively fix the pseudo-critical value of the coupling constant $\kappa_4^{\crit}(\bar{N}_{41})$ and  we study the properties of the model parametrized by the two remaining coupling constants: $\kappa_0$ and $\Delta$. Observing the geometric structures of individual triangulations and their extracted averaged properties, one may find regions where these properties differ from one another. In this way the two-dimensional phase diagram of the model can be determined, and it reveals a rich structure with many phase transition lines, see Fig. \ref{phasestructure}. The phase transitions in  CDT have so far been analyzed using  standard techniques. It has been shown that the peak in the variance of an order parameter typically designates, with high accuracy, the location of the phase transitions, and also the Binder cumulant can be used to predict the value of the infinite volume limit of a given coupling constant  using finite volume scaling measurements \cite{jordan, cb_1, cb_2, ac_1, bcbtrans1}. Typically, one selects a value of a coupling constant ($\kappa_0$ or $\Delta$) and creates a set of configurations with fixed values of one coupling (e.g., $\kappa_0$) and changed values of the other coupling (e.g., $\Delta$) with step-size $\delta \kappa_0$ or $\delta \Delta$. After thermalizing these configurations one collects statistics of various observables for different geometries. Collecting large enough statistics is necessary to calculate averages of these observables with high accuracy. Each line (or grid) of measurements gives information about the location of the phase transition and the pseudo-critical value of the observed quantities for fixed $\bar{N}_{41}$.\footnote{Note that in a lattice formulation for any finite lattice size $\bar{N}_{41}$ the free energy is finite and thus one has a cross-over rather than a true phase transition, thus we talk about \textit{pseudo-criticality}. Only in the limit where $\bar{N}_{41}\to \infty$ one recovers a true phase transition.}  Taking the limit $\bar{N}_{41} \to \infty$ gives the continuum behavior of the chosen observable, and the critical exponents can be extracted. The finite size dependence of the observables (which are functions of $\Delta$ and $\kappa_0$) is non-trivial. In this paper we try to use this dependence to determine the properties of the CDT phase transitions in the infinite volume limit, i.e., when $\bar{N}_{41}\to \infty$ and $\kappa_4 \to \kappa_4^{\crit}$.

One should also note that in the phase transition studies described herein, we have tested and used a new method of finding locations of the phase transition points based on machine learning techniques, namely the so-called {\it logistic regression} method which turned out to be very efficient.
More details can be found in Appendix 1.

\section{The phase diagram of CDT}\label{sec-phasediagram}

\begin{figure}[ht!]
\centering
\includegraphics[width=0.75\textwidth]{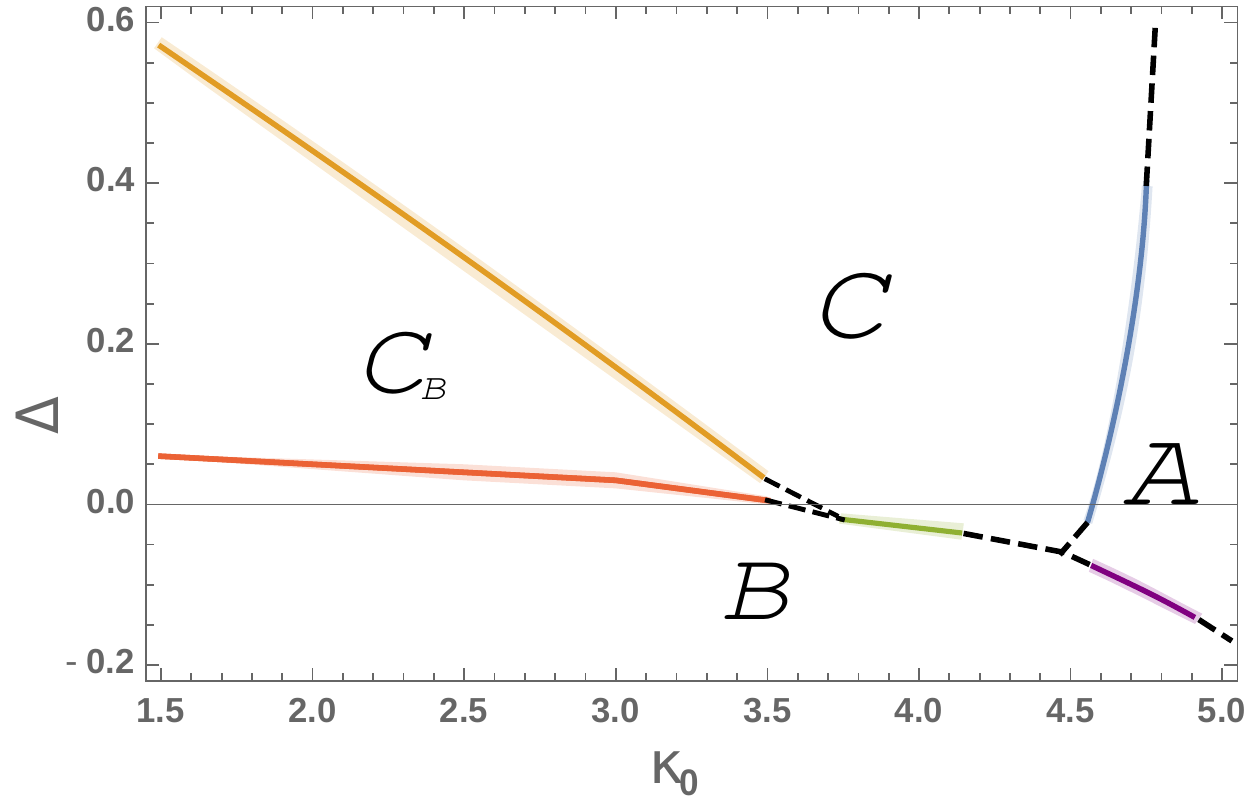}
\caption{Phase-diagram of CDT with the four phases. The plot shows the position of phase transitions measured for  fixed lattice volume $\bar N_{41} = 160\mathrm{k}$  in the toroidal CDT. In the spherical CDT case the phase structure is similar, although the phase transition between C and B phases (denoted by the green solid line) was inaccessible by numerical simulations.}
\label{phasestructure}
\end{figure}

\begin{figure}[ht!]
\centering
\frame{\includegraphics[width=0.23\textwidth]{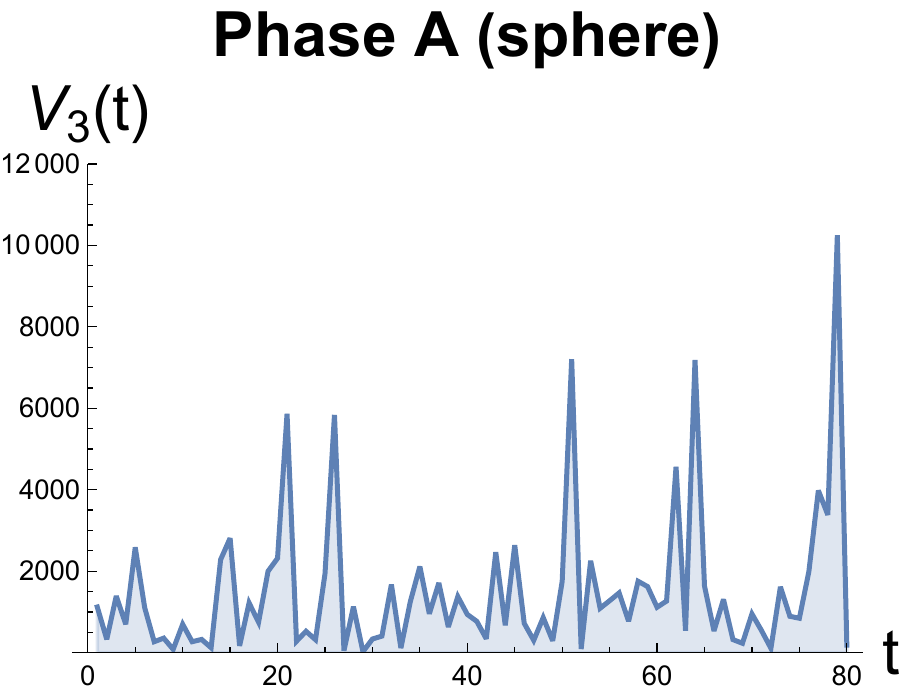}}
\frame{\includegraphics[width=0.23\textwidth]{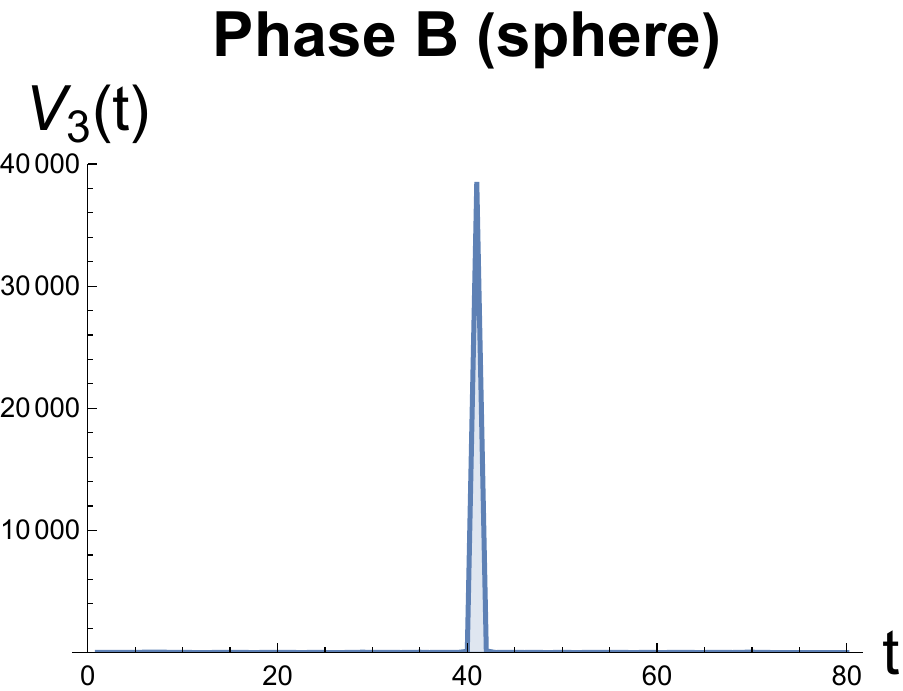}}
\frame{\includegraphics[width=0.23\textwidth]{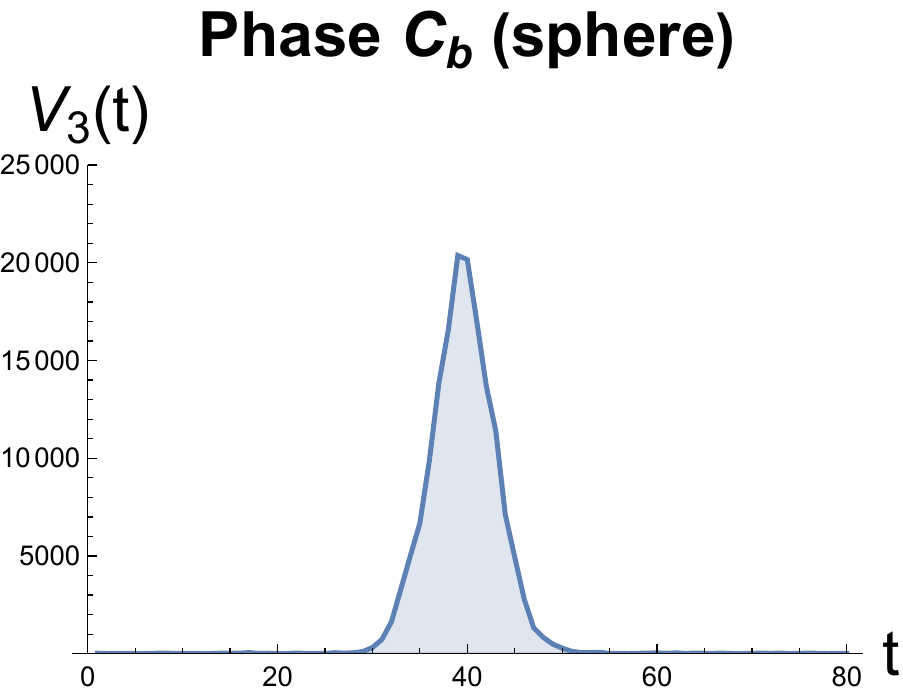}}
\frame{\includegraphics[width=0.23\textwidth]{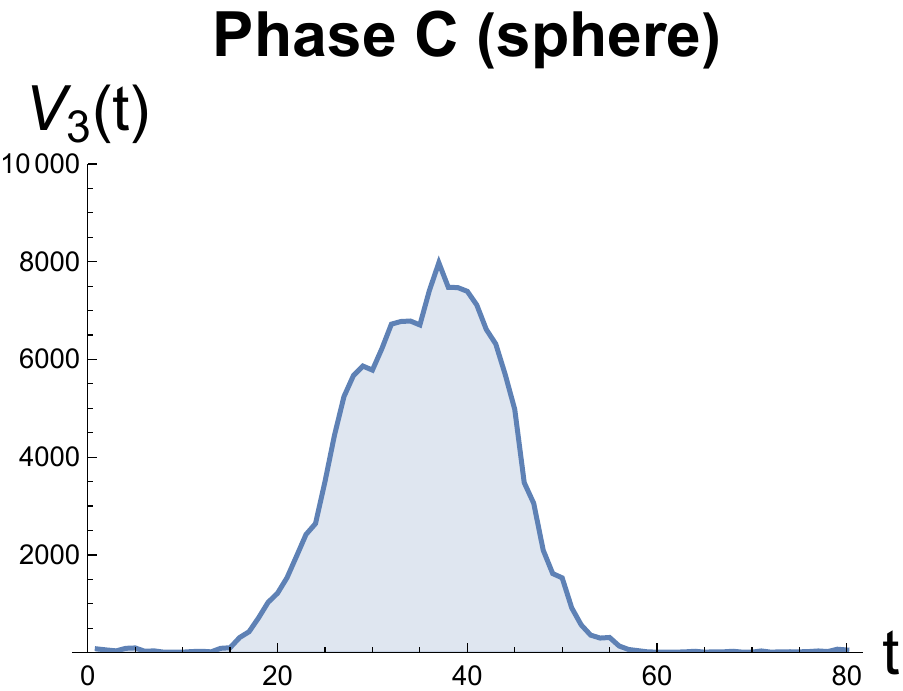}}\\
\frame{\includegraphics[width=0.23\textwidth]{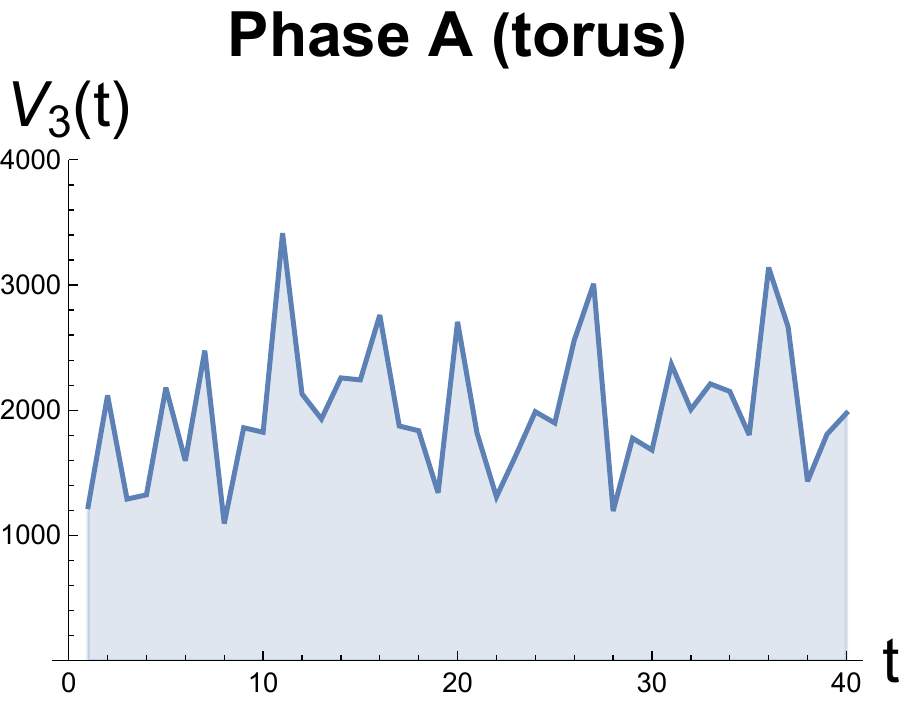}}
\frame{\includegraphics[width=0.23\textwidth]{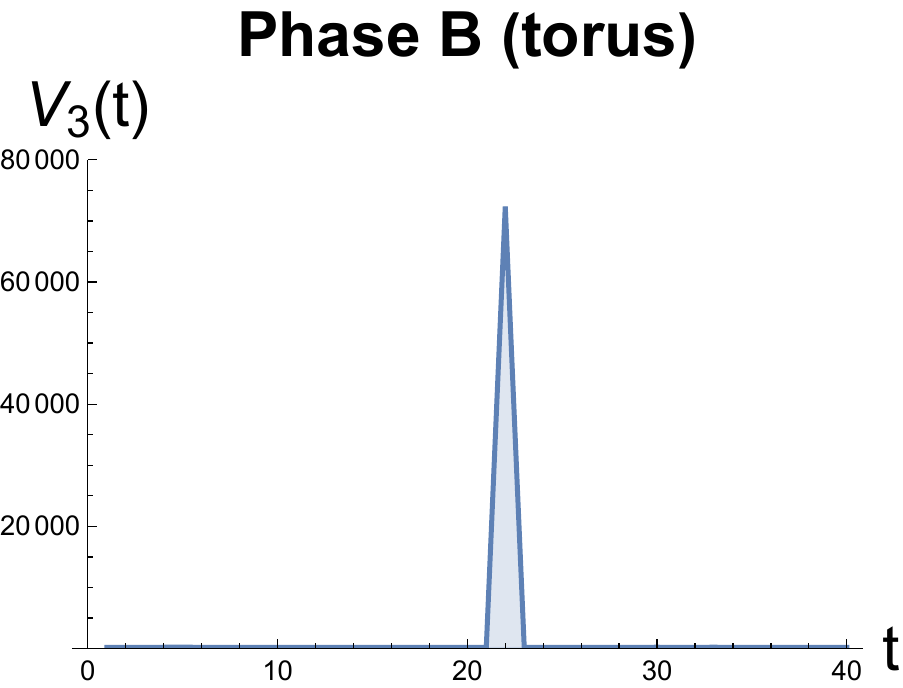}}
\frame{\includegraphics[width=0.23\textwidth]{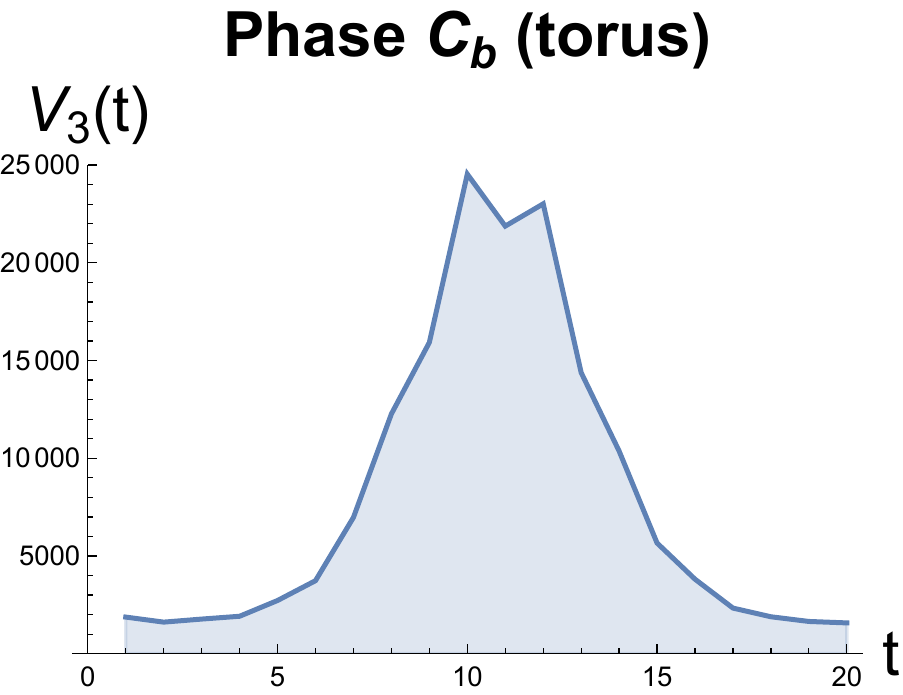}}
\frame{\includegraphics[width=0.23\textwidth]{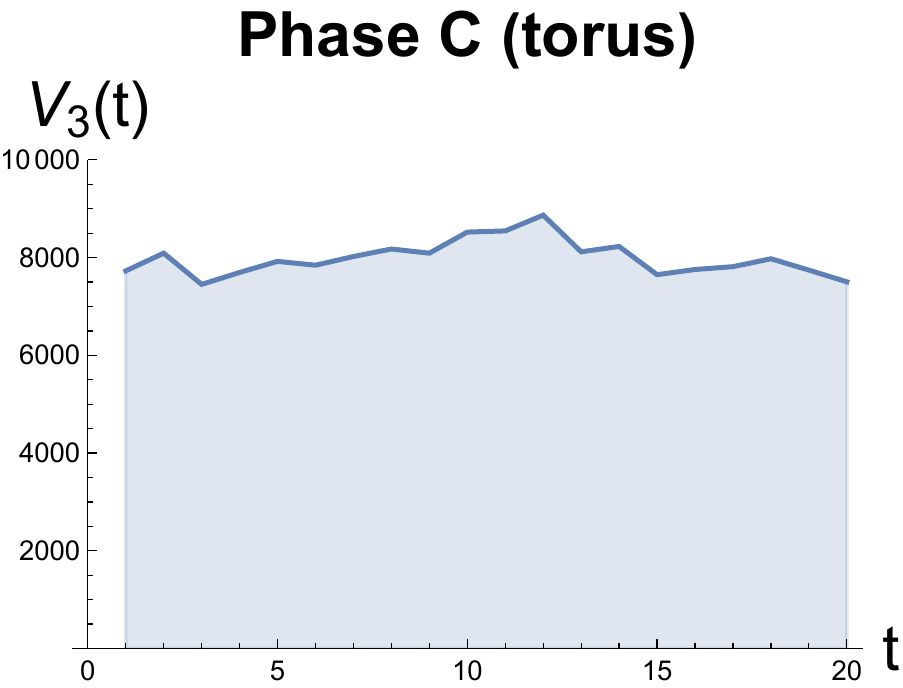}}\\
\caption{Spatial volume profiles of generic CDT configurations in different phases. Top: Spherical CDT: $A$, $B$, $C_b$, $C$; Bottom: Toroidal CDT:  $A$, $B$, $C_b$, $C$, respectively.}
\label{fig:volprofs}
\end{figure}

\begin{figure}[ht!]
\centering
\frame{\includegraphics[width=0.4\textwidth]{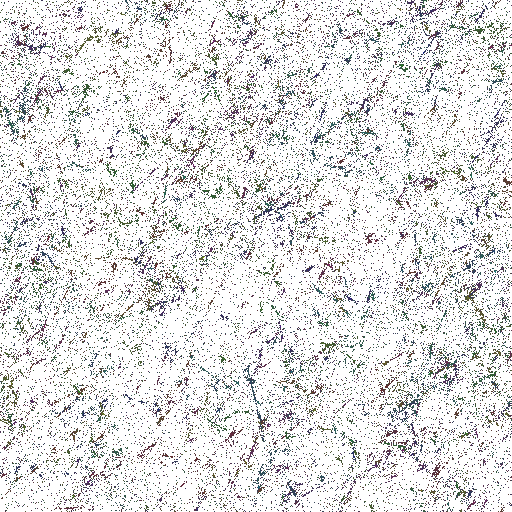}}
\frame{\includegraphics[width=0.4\textwidth]{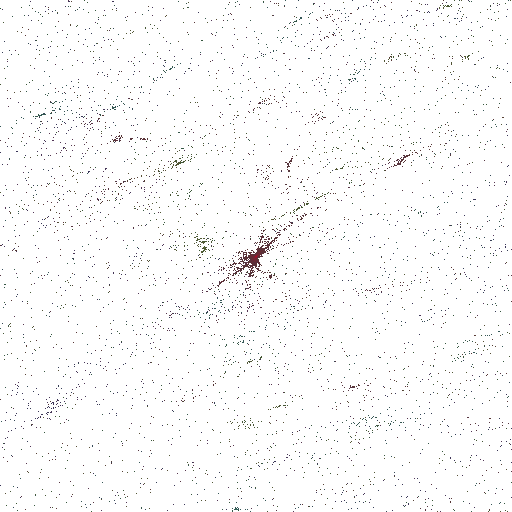}}\\
\frame{\includegraphics[width=0.4\textwidth]{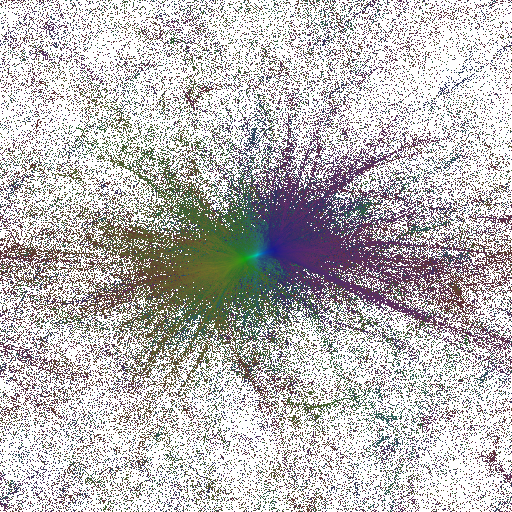}}
\frame{\includegraphics[width=0.4\textwidth]{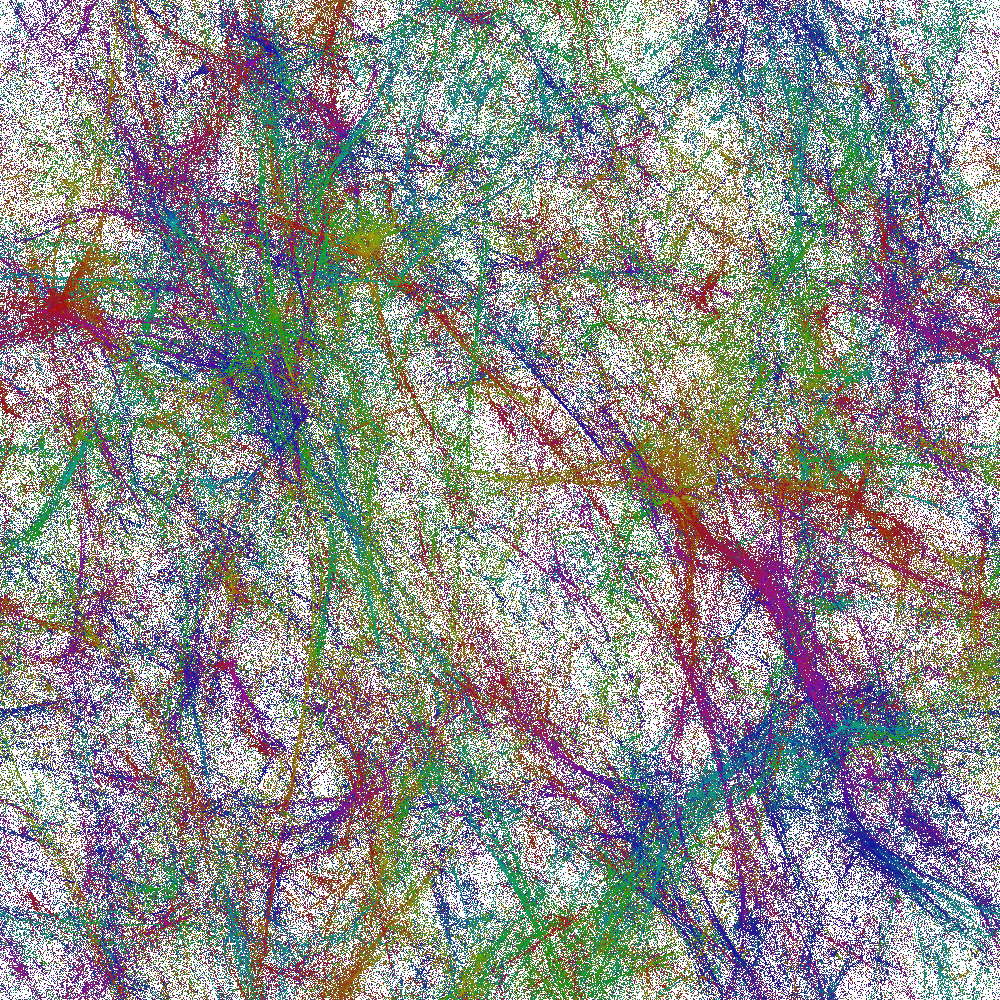}}
\caption{The structure of the geometry of phase $A$ (top left), $B$ (top right), $C_b$ (bottom left), and $C$ (bottom right), respectively, using the scalar field mapping method defined in \cite{scalar0}. The plots show four-volume density distribution of generic CDT configurations in each phase projected on two spatial directions.}
\label{fig:maps}
\end{figure}

The geometry of a CDT triangulation can be investigated by analyzing various {\it observables}.\footnote{Here what we call {\it observable} is a quantity which can be measured in numerical simulations. It does not have to be a gauge / diffeomorphism invariant observable.} They can be  general quantities characterizing global properties of a triangulation, e.g., the ratios: $\OP_1\equiv N_0/N_{41}$ or $\OP_2\equiv N_{32}/N_{41}$. They can be as well more local quantities, e.g., the distribution of the spatial three-volume as a function of the (lattice) time coordinate, which can be measured both in the spherical and the toroidal version of CDT, as it is presented in Fig. \ref{fig:volprofs}. Other examples of local observables are the recently introduced space-time coordinates \cite{scalar0}, which map a triangulation onto four scalar fields, enabling one to measure and visualize four-volume density distribution in all spacetime directions within the toroidal version of CDT, see Fig. \ref{fig:maps}.

Geometric properties of \textit{generic triangulations} dominating the path integral (\ref{PI}), and thus the typical behavior of the above-mentioned observables, will vary depending on the choice of the bare coupling constants: $\kappa_0$ and $\Delta$.\footnote{As explained above the $\kappa_4$ is fine tuned to the pseudo-critical value $\kappa_4^{\crit}(\bar N_{41})$ dependent on the fixed lattice volume $\bar N_{41}$.} Even though the bare  action of CDT (\ref{Action}) is very simple, the $3+1$-dimensional model reveals a surprisingly rich phase structure, see Fig.~\ref{phasestructure}, which seems to be  universal, independent of the spatial  topology choice \cite{phase_diag}. So far, four phases with distinct geometric properties were discovered - called $A$, $B$, $C$ and $C_b$, respectively.

For large values of the bare inverse gravitational constant $\kappa_0$ one encounters phase $A$. This phase is characterized by the spatial volumes at different (lattice) time layers being uncorrelated \cite{transfer_m}, which is seen clearly in the upper left panel in Fig.\ \ref{fig:volprofs}. Approaching phase~$A$ from the phase $C$, where one has an effective minisuperspace action \cite{hartle_hawking} (see below), the coefficient in front of the kinetic term, which couples spatial volume in the neighboring time layers, vanishes reflecting the lack of correlation between spatial volumes. On the $A$-side the system is described by an effective action without kinetic term. On the other hand, the generic geometry of the spatial slices themselves seems to be isotropic and homogeneous, see Fig.\ \ref{fig:maps}, left upper panel. 

For small or negative values of the asymmetry parameter $\Delta$ one observes phase $B$, where the typical spatial volume distribution is collapsed into a single time layer with the three-volume distribution pinched to the cut-off size in the remaining part of a generic triangulation, see Fig. \ref{fig:volprofs}. Thus, the four-dimensional geometry is effectively reduced to a three-dimensional geometry, in the sense that all four-volume is contained in a thin slab surrounding a single time layer. Moreover, in such a generic configuration, the isotropy and homogeneity of three-dimensional space is itself maximally broken, in the sense that the whole volume is concentrated around a few vertices. This is illustrated in Fig.\ \ref{fig:maps}, top right panel.

For not too large $\kappa_0$, increasing $\Delta$ one moves from the $B$ phase to the so-called bifurcation phase, $C_b$. In this phase the number of constant time layers where the spatial volume is significantly different from the minimal value starts to increase and forms a \textit{blob}, see Fig. \ref{fig:volprofs}. However, when one increases the four-volume $N_{41}$ sufficiently, the extension of this blob ceases to increase as $N_{41}^{1/4}$. This implies that in the limit where $N_{41} \to \infty$, the \textit{stalk}, where the spatial volumes are of cut-off scale, will dominate if we allow the time direction to be of order at least $N^{1/4}_{41}$. In addition, if we look at the spatial geometry inside time layers belonging to the blob, it shows the same characteristics as the spatial geometry in the $B$ phase: isotropy and homogeneity are broken for a generic configuration by the presence of a few high-order vertices. These high-order vertices are correlated from slice to slice within the blob. As we continue to increase $\Delta$ or $\kappa_0$ this effect becomes less and less pronounced \cite{nilas}: the order of the high-order vertices decreases and the blob broadens but it does not disappear. A stalk will remain. The lower left panel in Fig.\ \ref{fig:maps} shows that not only in the time direction, but also in the spacial directions, the geometry is more extended than in the $B$~phase, but still inhomogeneous. Increasing $\Delta$ or $\kappa_0$ even further bring us into phase~$C$.

In phase $C$, which is also denoted the de Sitter phase or the semiclassical phase, the spatial volume profiles seem to depend on the choice of spatial topology, as illustrated in the right panels in Fig.\  \ref{fig:volprofs}. In the case where the spatial topology is $S^3$ the volume profiles form a blob, like in the bifurcation phase. The crucial difference is that the extension of the blob scales as $N_{41}^{1/4}$ and a typical spatial volume in the blob scales as $N_{41}^{3/4}$. Thus, one might obtain a semiclassical geometry of the blob in the limit $N_{41} \to \infty$. Similarly, the generic spatial slices seem to be isotropic and homogeneous without the high-order vertices of phase $B$ and phase $C_b$. In the case where the spatial topology is $T^3$ we do not observe a blob, but rather a constant volume profile with superimposed (small) fluctuations. In both cases, the spatial geometries (except for the geometry in the stalk in the case where the spatial topology is $S^3$) of generic configurations seem to be isotropic and homogeneous on large scales while on shorter scales one can observe volume density fluctuations forming voids-and-filaments \cite{scalar0, scalar2}, surprisingly similar to structures formed by matter content of the real Universe, as illustrated in the lower right panel in  Fig.\ \ref{fig:maps}. In accordance with this, the volume profiles are well described by minisuperspace actions similar to the Hartle-Hawking minisuperspace action \cite{deSitter, semiclassical, transfer_1, impactoftopology, transfer_2}, and the kinetic term in the minisuperspace actions has the same coefficient in the two cases.

The $A\mi C$ transition has been determined to be a first-order transition \cite{jordan, ac_1}, while the  the $B\mi C_b$ transition has been determined to be a second-order transition \cite{jordan, bcbtrans1} in both $S^3$ and $T^3$ spatial topology choices. The $C\mi C_b$ transition was found to be a second-order transition in the case where the spatial topology was $S^3$ \cite{cb_1, cb_2} and  (seemingly) a first-order transition when the spatial topology was $T^3$ \cite{phase_diag}. The $B\mi C$ transition and the $A \mi B$ transition could be studied by MC simulations in the case where the spatial topology is $T^3$. Preliminary results for the $B\mi C$ transition were reported in \cite{towardsUV}, but herein we will improve the data for the $B \mi C$ transition and complete the phase diagram by measuring the yet unexplored $A-B$ transition.  As a result, we will see an 
interesting pattern emerge, to be discussed in Sec.\ \ref{discussion}.

\section{The $A \mi B$ phase transition}\label{sec:AB}

The phase transition between phases $A$ and $B$ was not analyzed earlier in detail. This transition is important because there is a point in the phase diagram which is the common endpoint of the $A \mi C$, $A\mi B$ and $B \mi C$ phase transitions (the $A\mi B \mi C$ triple point), see Fig. \ref{phasestructure}, which is a potential candidate of the UV fixed point of CDT. Thus, understanding the nature of all surrounding phase transitions is crucial. We perform a finite volume scaling analysis and determine the order of the $A \mi B$ phase transition for three different fixed values of $\kappa_0 \equ 4.8$, 4.6 and 4.5 (and the corresponding observables will be denoted with their corresponding $\kappa_0$ coupling as a lower right index). At this phase transition, the time-reduced collapsed configurations on the $B$ side of the transition compete with time-uncorrelated configurations on the $A$ side (see Fig. \ref{fig:volprofs}). For consistency, the scaling exponent is measured for the three fixed values of $\kappa_0$ independently by varying the values of $\Delta$ to find the (pseudo-)critical values $\Delta^{\crit}(\bar{N}_{41})$ together with the scaling exponents $\gamma$. We performed simulations for various volumes (ranging between $\bar N_{41} \equ 20\mathrm{k}$ and $720\mathrm{k}$) to find the scaling exponents for each fixed $\kappa_0$ measurement set. Our assumption about the critical behavior of $\Delta^{\crit}(\bar{N}_{41})$ in the limit $\bar N_{41}\to \infty$ is a (standard) power function, the same as used in the previous phase transition studies cited before:
\begin{equation}
\label{eq_scaling0}
\Delta^{\crit}(\bar{N}_{41}) = \Delta^{\infty} - \tilde{C} \cdot \bar{N}_{41}^{-1/\gamma},
\end{equation}
where $\tilde{C}$ is a constant, $\Delta^{\infty}$ is the  value of the asymmetry parameter corresponding to the infinite volume limit at fixed $\kappa_0$ and $\gamma$ is the scaling exponent. 

\begin{figure}[h!]
\includegraphics[width=0.8\textwidth]{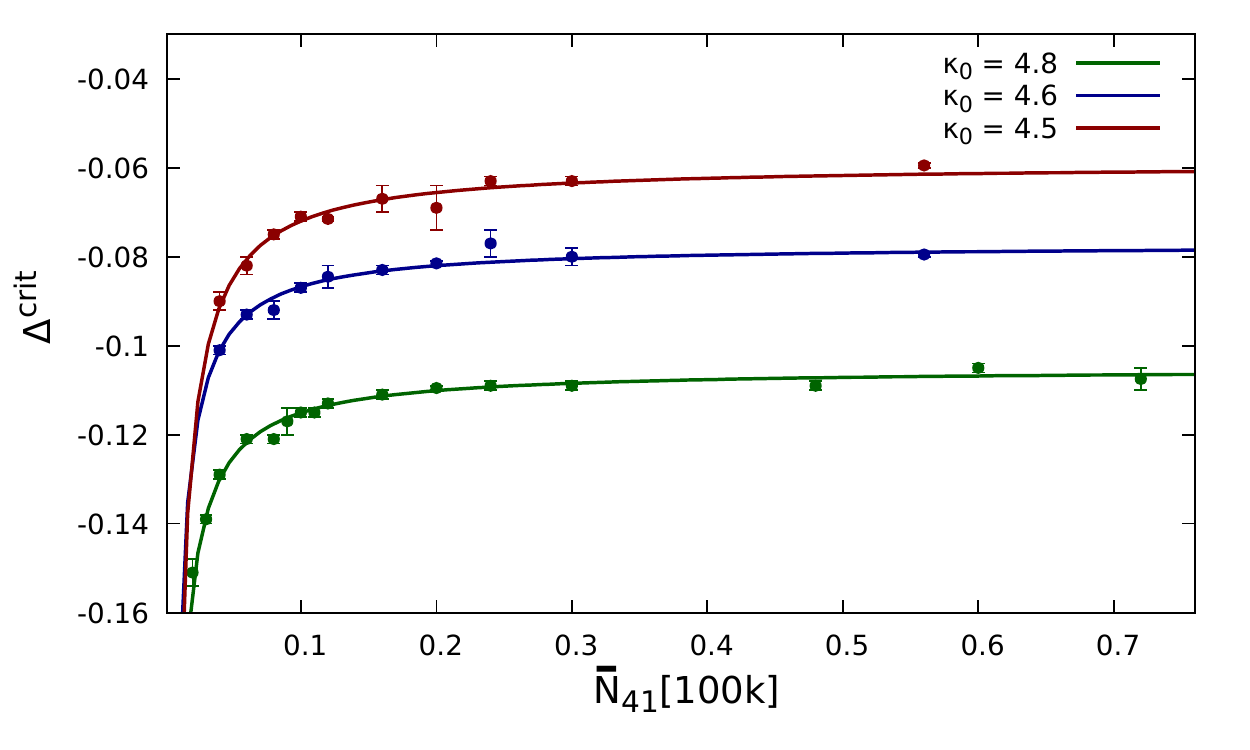}
\centering
\caption{(Pseudo-)critical values of $\Delta^{\crit}(\bar N_{41})$ measured for  $\kappa_0 \equ 4.8$ (green), $\kappa_0 \equ 4.6$ (blue), and $\kappa_0 \equ 4.5$ (red) together with the fits of eq. (\ref{eq_scaling0}). The solid curves  were fitted with the critical exponent fixed to $\gamma \equ 1$ for all  three data sets.}
\label{fig_ab_delta}
\end{figure}

The best fits for the scaling exponents $\gamma$ were measured to be $\gamma_{4.8} \equ 1.088 \pm 0.101$, $\gamma_{4.6} \equ 1.029 \pm 0.178$, and $\gamma_{4.5} \equ 1.151 \pm 0.379$ for the three $\kappa_0$ values considered. As the best fitted exponent values are all consistent with the first-order behavior, i.e., $\gamma \equ 1$, we used fixed $\gamma = 1$ to fit relation (\ref{eq_scaling0}) to our data, as presented in Fig. \ref{fig_ab_delta}. By extrapolating the fits to the infinite volume limits $\Delta^{\crit}$ were determined to be $\Delta^{\infty}_{4.8} \equ -0.110 \pm 0.001$,  $\Delta^{\infty}_{4.6} \equ -0.077 \pm 0.001$ and $\Delta^{\infty}_{4.5} \equ -0.059 \pm 0.001 $ for the three cases respectively.\\

\begin{figure}[ht!]
\centering
\includegraphics[width = 0.8\textwidth]{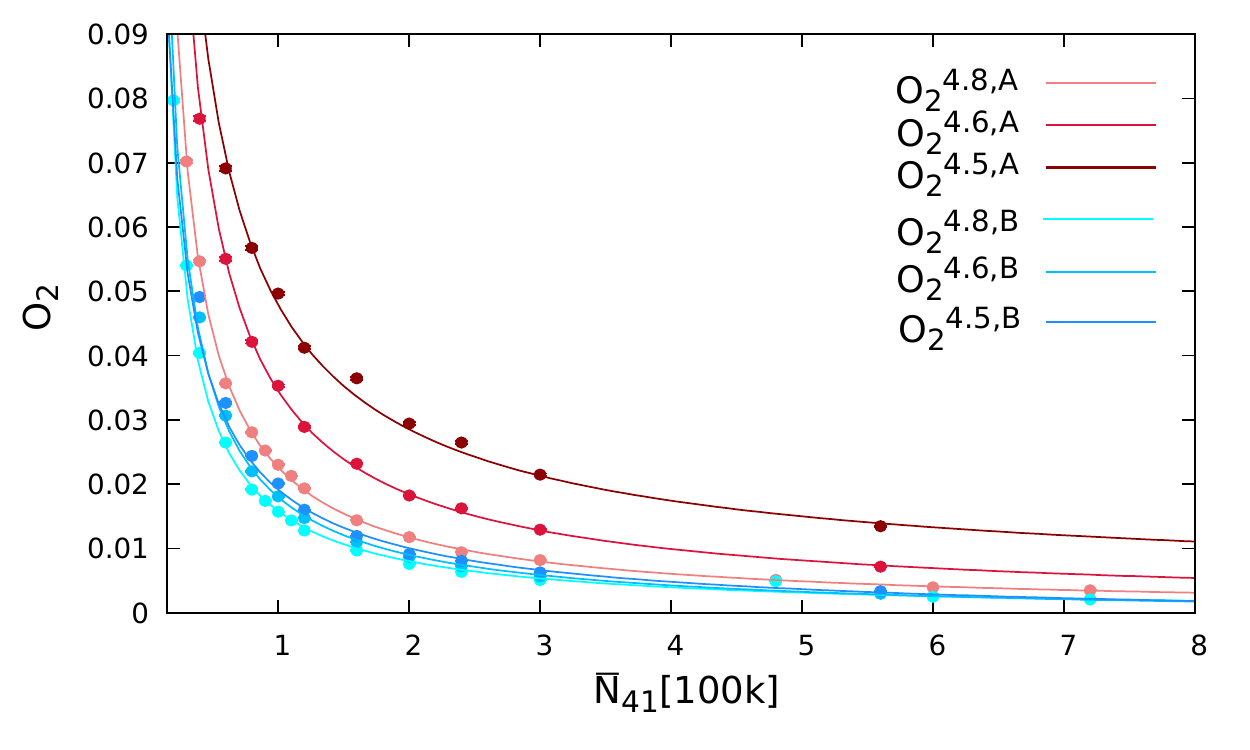}
\caption{The running of $\mathcal{O}_2$ for $\kappa_0 = 4.8$, $\kappa_0 = 4.6$ and $\kappa_0 = 4.5$.  Blue colors correspond to data measured in phase $B$ and red in phase $A$ closest to the phase transition point, and the darker the color the lower the corresponding $\kappa_0$ coupling, i.e., the closer to the $A\mi B \mi C$ triple point. The error bars are smaller than the size of the data-points. The solid curves correspond to the fits of a relation similar to eq. (\ref{eq_scaling0}) with the critical exponent fixed to be $\gamma = 1$ for all  data sets. }
\label{fig:ab_op2}
\end{figure}

In the following analysis, we will assume that near the phase transition point not only the critical coupling(s)  $\Delta^{\crit}(\bar N_{41})$, but also other relevant order parameters scale similarly to eq. (\ref{eq_scaling0}) with the critical exponent fixed at $\gamma =1$. This is indeed supported by our data. In Fig. \ref{fig:ab_op2}  we show the results for the $\mathcal{O}_2\equiv N_{32}/N_{41}$ observable measured for each side of the phase transition independently.

 The fits shown in Fig. \ref{fig:ab_op2} have a close to zero value of $\mathcal{O}_2^\infty$ for all cases, but on the $A$ side of the transition the approach to $\mathcal{O}_2^\infty \approx 0$
becomes slower with decreasing $\kappa_0$, i.e., slower when we approach the $A\mi B\mi C$ triple point.

\begin{figure}[ht!]
\centering
\includegraphics[width=0.8\textwidth]{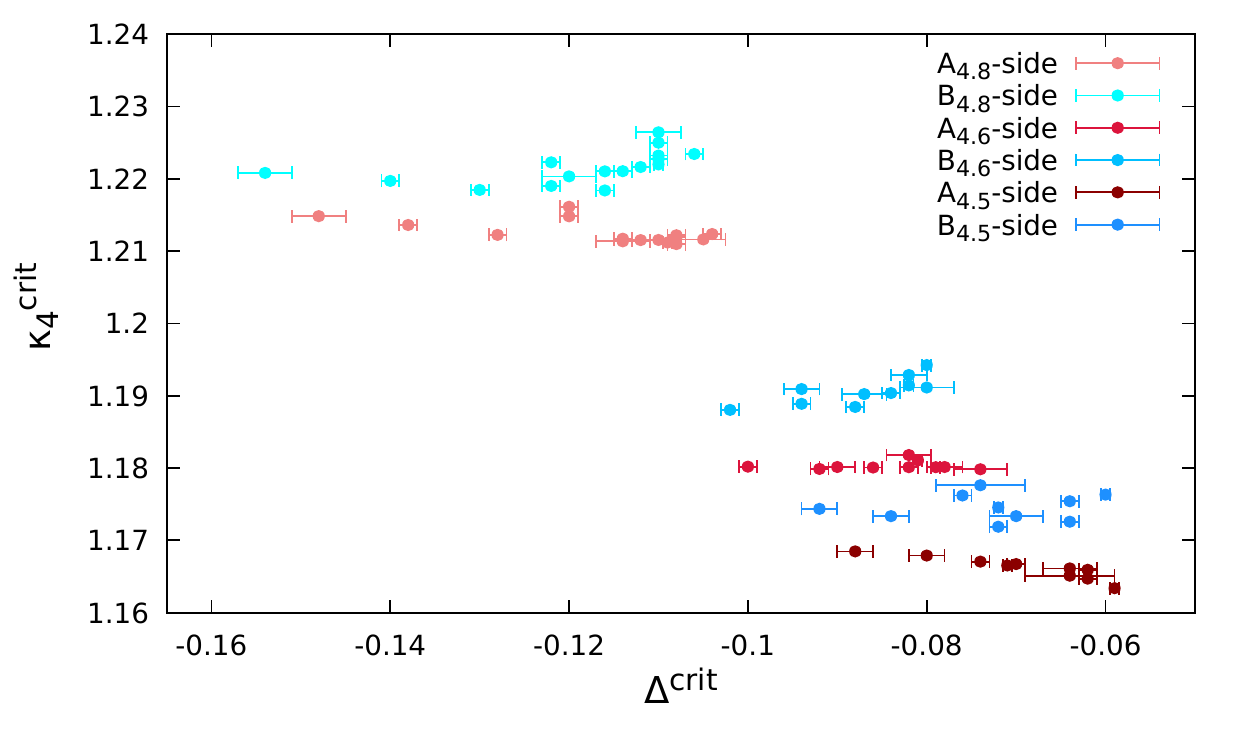}
\caption{Values of the critical  $\kappa_4^{\crit}(\bar N_{41})$ coupling measured as a function of $\Delta^{\crit}(\bar N_{41})$  in phase $B$ (red) and phase $A$ (blue) closest to the phase transition points.  The darker the color the lower the corresponding $\kappa_0$ coupling, i.e., the closer to the $A\mi B\mi C$ triple point. Data suggest the lack of a common value of $\kappa_4^{\infty}$ for the $A$ and $B$ phases at the infinite volume limit.}
\label{fig_ab_k4}
\end{figure}

A final observation corroborating the first-order nature of the $A \mi B$ transition is the fact that the values $\kappa_4^{\crit}(\bar{N}_{41})$ observed in the $A$ phase  and the $B$ phase data closest to the transition point do not converge to a common value for $\bar{N}_{41} \to \infty$. This is shown in Fig.\ \ref{fig_ab_k4} (where we have used data from Fig.\ \ref{fig_ab_delta} to express $\bar{N}_{41}$ in terms of $\Delta^{\crit}$). It is seen that when we increase $\bar{N}_{41}$, which corresponds to higher $\Delta^{\crit}$ in the plot, the gap between $\kappa_4^{\crit}(\mbox{A- side})$ and $\kappa_4^{\crit}( \mbox{B- side})$ increases. This shows that the MC simulations collect very different configurations on the $A$ and the $B$ phase transition side, a situation typical for a first-order transition.\\

Summarizing, in this section we have analyzed the behavior of the $A\mi B$ phase transition for three different fixed values of $\kappa_0$ via the scaling of the asymmetry parameter $\Delta^{\crit}(\bar{N}_{41})$ and $\mathcal{O}_2$ on the two sides of the phase transition. For all measured values of $\kappa_0$, the fitted values of the scaling exponent $\gamma$ are consistent with the first-order behavior ($\gamma \equ 1$). Thus, we conclude that the $A\mi B$ phase transition is a first-order phase transition. The order of the $A \mi B$ phase transition does not exclude the possibility that exactly at the endpoint ($A\mi B\mi C$ triple point) the phase transition could be of higher-order.

\section{Revisiting the $B \mi C$ phase transition}\label{sec:BC}

Recently, using manifolds with toroidal spatial topology, we were able, for the first time, to measure the properties of the $B\mi C$ phase transition \cite{towardsUV} at fixed $\kappa_0 \equ 4.0$. This result was  important since this region of the phase diagram is between the two common endpoints of CDT phase transition lines (the $A\mi B\mi C$ and the $C_b\mi B\mi C$ triple points), see Fig. \ref{phasestructure}. We determined the critical behavior based on measurements for a sequence of values of $\Delta$ approaching the phase transition from $B$ and $C$ sides of the phase diagram. Using the approach presented in \cite{towardsUV} we established a range of coupling constants, within which, for a given finite value of $\bar{N}_{41}$, the system jumps between the two phases. We determined the limits of this hysteresis region and measured the (volume dependent) pseudo-critical values of the coupling constants $\Delta$ and $\kappa_4$ as close to the phase transition as we could get with respect to the hysteresis. We observed that, similarly to the $A \mi B$ transition case, the position of the phase transition line moves towards larger $\Delta$ and $\kappa_0$ values when $\bar{N}_{41}$ is increased. Again, the true phase transition is observed in the infinite volume limit $\bar{N}_{41}\to \infty$ which corresponds to $\kappa_4 \to \kappa_4^{\crit}(\kappa_0,\Delta)$. In \cite{towardsUV} we assumed the standard finite volume scaling for $\Delta^{\crit}(\bar{N}_{41})$, $\kappa_0^{\crit}(\bar{N}_{41})$ and $\kappa_4^{\crit}(\bar{N}_{41})$, analogous to eq. (\ref{eq_scaling0}): 
\begin{eqnarray}
\label{eq_scaling}
\Delta^{\crit}(\bar{N}_{41}) &=& \Delta^{\infty} - A \cdot \bar{N}_{41}^{-{1}/{\gamma}}\\ \nonumber
\kappa_0^{\crit}(\bar{N}_{41}) &=& \kappa_0^{\infty} - B \cdot \bar{N}_{41}^{-{1}/{\gamma'}}\\ \nonumber
\kappa_4^{\crit}(\bar{N}_{41}) &=&\kappa_4^{\infty}- C \cdot \bar{N}_{41}^{-{1}/{\gamma''}}.
\end{eqnarray}
We showed that in the limit $\bar{N}_{41}\to \infty$, the hysteresis shrinks to zero.
Furthermore, we observed that on both sides of the phase transition $\Delta^{\crit}(\bar{N}_{41})$ and $\kappa_4^{\crit}(\bar{N}_{41})$ scale in the same way. The scaling analysis of this behavior could be characterized by a common critical exponent $\gamma \approx 1.62 \pm 0.25$, potentially signaling a higher-order phase transition. As we will show below, an equally good fit to the data can be obtained using $\gamma \equ 1$ and a higher-order correction given by a universal small constant shift of $N_{41}$ for data measured on both sides of the transition line. This  indicates that the transition might be first-order, which is supported by the analysis of the order parameters $\OP_1\equiv N_0/N_{41}$ and $\OP_2 \equiv N_{32} / N_{41}$ showing a strong discontinuity between the two phases even in the infinite volume limit, as already noted in \cite{towardsUV}. 

\begin{figure}[ht!]
\centering
\includegraphics[width = 0.45\textwidth]{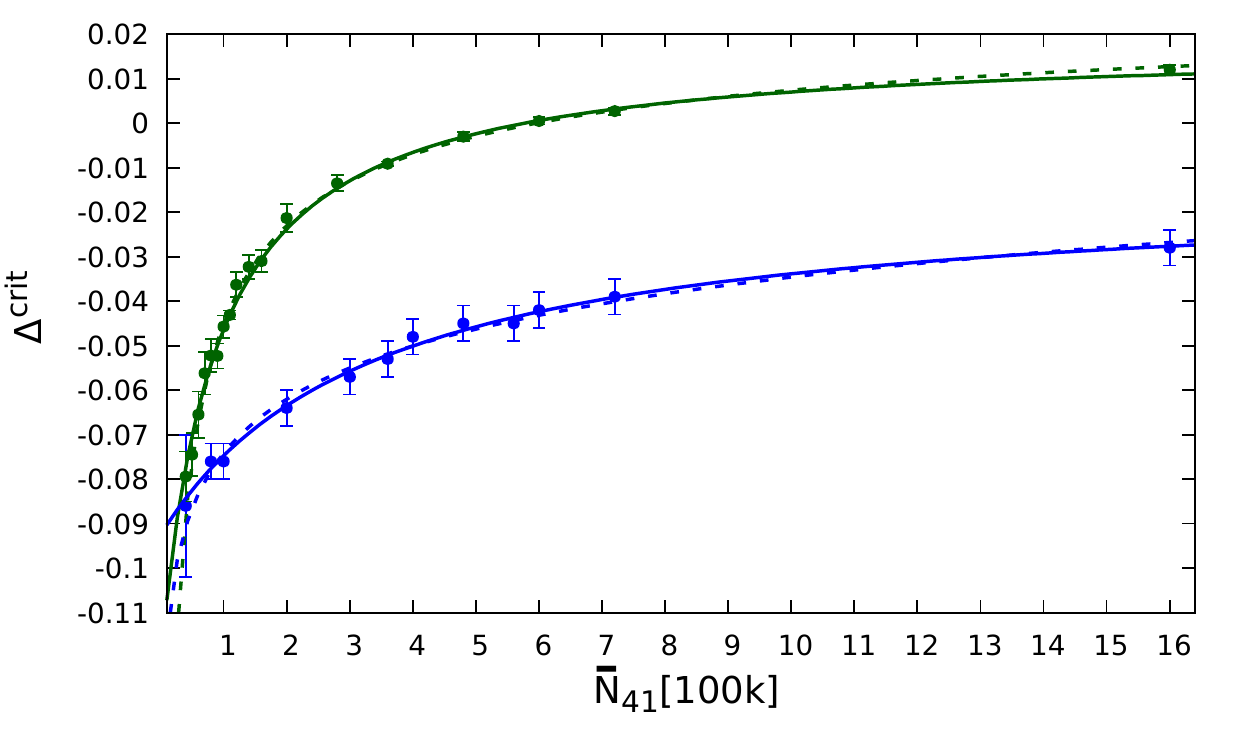}
\includegraphics[width = 0.45\textwidth]{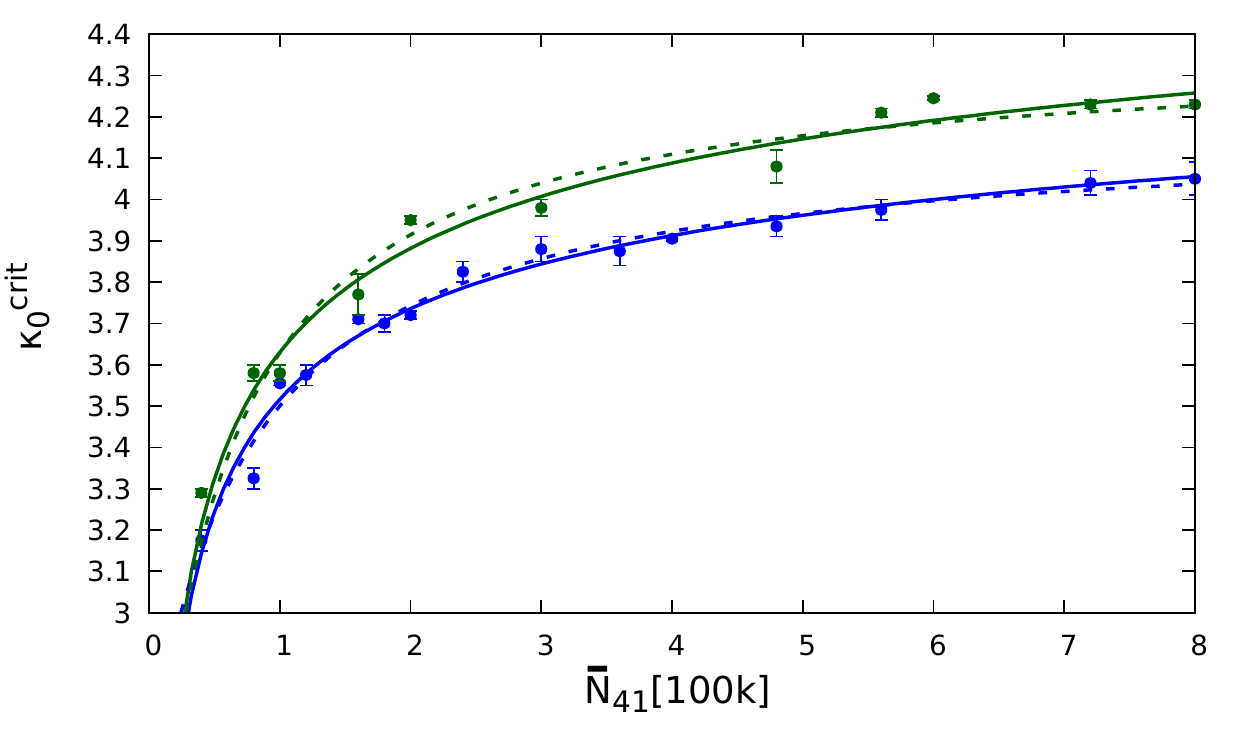}
\caption{Finite volume scaling of the coupling constants $\Delta^{\crit}$ (left panel) and $\kappa_0^{\crit}$ (right panel). The dashed and solid curves represent fits to eq. (\ref{eq_scaling}) and eq. (\ref{eq_scaling_m1}), respectively. In the left panel (vertical measurements), green data-points are for fixed $\kappa_0 = 4.0$ and blue are for $\kappa_0 = 4.2$. In the right panel (horizontal measurements), green data-points are for fixed $\Delta = -0.02$ and  blue are for $\Delta = 0$. }
\label{fig:scaling_couplings}
\end{figure}

Herein we have repeated the analysis for $\kappa_0 = 4.0$ with much larger statistics and we have also performed new measurements for $\kappa_0 = 4.2$, which is closer to the $A\mi B\mi C$ triple point. The lattice volumes used ranged from $\bar{N}_{41} = 40\mathrm{k}$ and $1600\mathrm{k}$. We denote the measurements performed at fixed $\kappa_0 = 4.0$ and $4.2$ as \textit{vertical} since in the $(\kappa_0,\Delta)$ coupling-constant plane only $\Delta$ varies. Similarly, we have performed measurements with fixed $\Delta=0$ and $-0.02$ varying $\kappa_0$ values. We refer to these as \textit{horizontal} measurements. In the horizontal measurements, the lattice volumes ranged from $\bar{N}_{41} = 40\mathrm{k}$ to $800\mathrm{k}$. n the following, we will present results in pairs related to the two values of the fixed coupling constant in the horizontal and the vertical measurement, respectively. Even though we now have much better statistics than in \cite{towardsUV} and have measured the $B\mi C$ phase transition at various bare coupling constant locations, approaching the transition line in different ways (vertically and horizontally), a fit to eq. (\ref{eq_scaling}) still does not allow a very good determination of the scaling exponents $\gamma,\gamma'$ and $\gamma''$, indicating that for the considered range of lattice volumes the assumed functional form (\ref{eq_scaling}) might not be optimal. Typically, we could only determine $\gamma$ to be between $1.6$ and $2.6$, but on the other hand $\gamma =1$ does not seem to be a good fit. This leads to the conclusion that higher-order corrections should be included as they can be important for smaller volumes. Below we propose a very simple finite size correction of the kind considered in \cite{ajw}, namely we perform a shift $\bar{N}_{41} \to \bar{N}_{41} - \mathrm{const}$. Thus, instead of using eq. (\ref{eq_scaling}), we will use a slightly modified  scaling relation, which can be viewed as corresponding to the first-order transition (fixed $\gamma \equ 1$),
but with a specific form of a finite size correction:
\begin{eqnarray}
\label{eq_scaling_m1}
\Delta^{\crit}(\bar{N}_{41}) &=& \Delta^{\infty} - 
A \cdot (\bar{N}_{41} - \mathrm{const}_1)^{-1},\\ \nonumber
\kappa_0^{\crit}(\bar{N}_{41}) &=& \kappa_0^{\infty} - 
B \cdot (\bar{N}_{41} - \mathrm{const}_2)^{-1},\\ \nonumber
\kappa_4^{\crit}(\bar{N}_{41}) &=&\kappa_4^{\infty}- 
C \cdot (\bar{N}_{41} - \mathrm{const}_3)^{-1}.
\end{eqnarray}
Both scaling relations of eq. (\ref{eq_scaling}) and eq.\ (\ref{eq_scaling_m1}) have the same number of free parameters, i.e., they require a three-parameter fit.
We can now compare the two different classes of fits. The parameters for the best fits of the critical scaling of $\Delta^{\crit}$ (vertical measurements) and $\kappa^{\crit}$ (horizontal measurements) using equations (\ref{eq_scaling}) and (\ref{eq_scaling_m1}) were found via the least squares method. In Table \ref{table:1} we show values of $\chi^2$ corresponding to the fits presented in Fig.\ \ref{fig:scaling_couplings}.
\begin{table}[h!]
\centering
\begin{tabular}{||c| c c| c c||} 
 \hline
 Fit & $\chi^2_{\kappa_0 = 4.0}$ & $\chi^2_{\kappa_0 = 4.2}$ & $\chi^2_{\Delta = 0}$ & $\chi^2_{\Delta = -0.02}$ \\ [0.5ex] 
 \hline\hline
 free $\gamma$ & 1.3e-04& 6.4e-05 & 2.3e-02 & 3.6e-02  \\
 $\gamma = 1$ & 1.7e-05 & 4.6e-05 & 1.9e-02 & 4.7e-02 \\ [1ex]
 \hline
\end{tabular}
\caption{The $\chi^2$ values related to the individual fits of  eq. (\ref{eq_scaling}) (free $\gamma$) and  eq. (\ref{eq_scaling_m1}) ($\gamma = 1$) to finite volume scaling of the coupling constants: $\Delta^{\crit}$ (vertical measurements in the left two columns) and $\kappa_0^{\crit}$ (horizontal measurements in the right two columns).}
\label{table:1}
\end{table}

\begin{figure}[ht!]
\centering
\includegraphics[width = 0.45\textwidth]{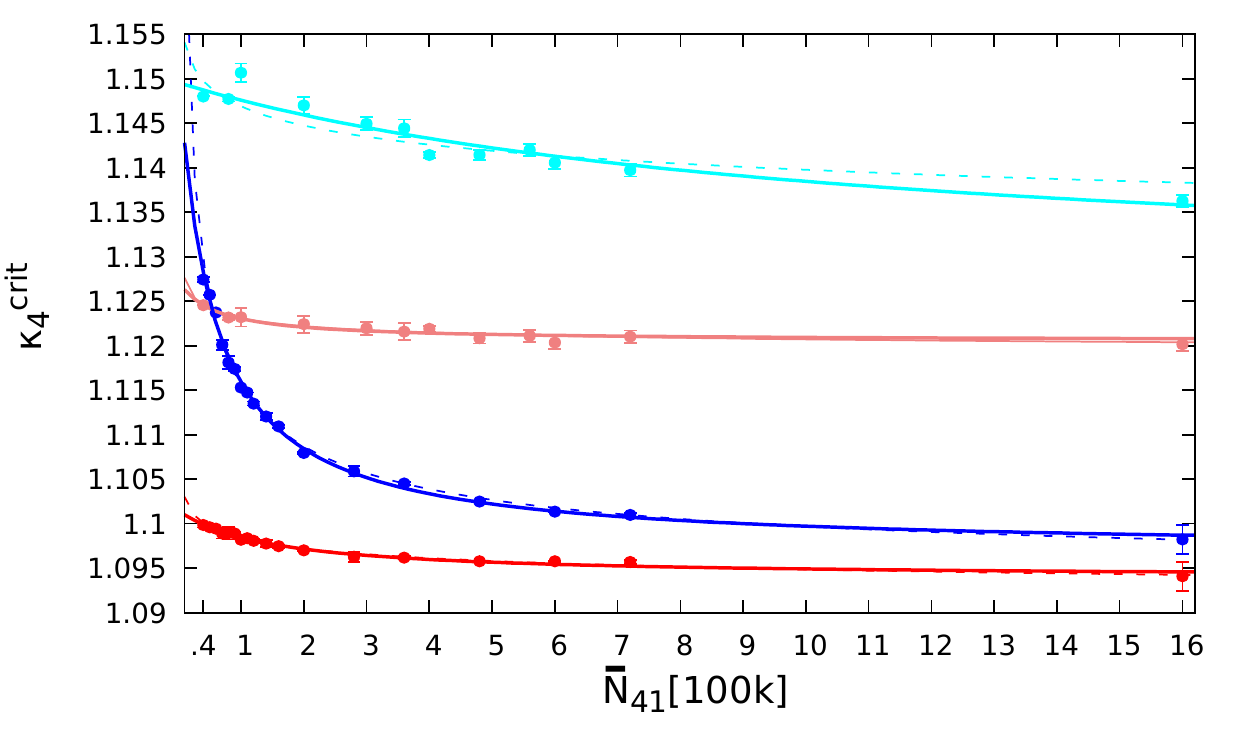}
\includegraphics[width = 0.45\textwidth]{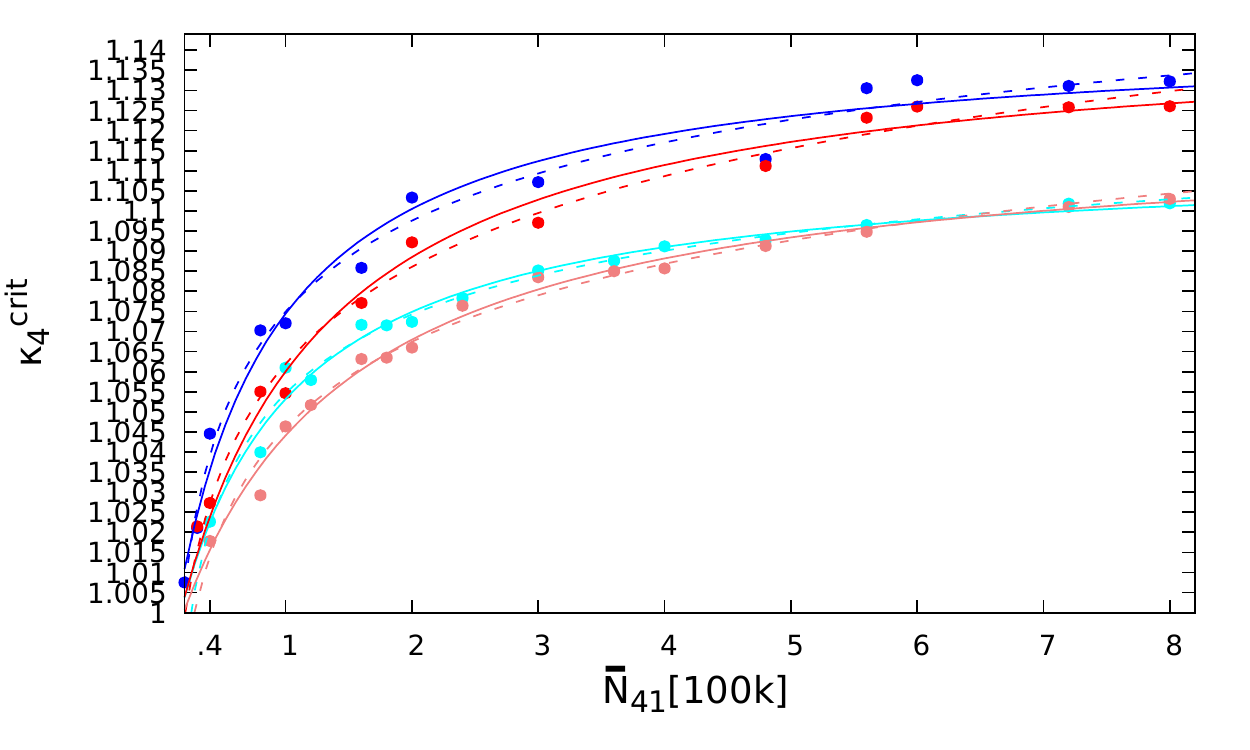}
\caption{Finite volume scaling of the $\kappa_4^{\crit}$ in vertical measurements with fixed $\kappa_0=4.2$ and $4.0$ (left panel) and horizontal measurements with fixed $\Delta=0$ and $-0.02$ (right panel). The dashed and solid curves represent fits to eq. (\ref{eq_scaling}) and eq. (\ref{eq_scaling_m1}), respectively. In both plots blue color corresponds to data measured in phase $B$ and red in phase $C$ - closest to the phase transition (hysteresis) region. Darker colors correspond to lower values of the fixed coupling constant.}

\label{fig:k4_crit}
\end{figure}

In Fig.\ \ref{fig:scaling_couplings} the vertical measurements (for fixed $\kappa_0$) are in the left plot and the horizontal ones (for fixed $\Delta$) are in the right plot. 
In both cases, the dashed lines show the fits of eq. (\ref{eq_scaling}) and solid lines are the fits of eq. (\ref{eq_scaling_m1}). Table \ref{table:1} and Fig. \ref{fig:scaling_couplings} clearly show that the two classes of fits are really close to each other in all cases; thus, using this information alone, it is hard to distinguish the order of the $B \mi C$ transition. Additional hint comes from  analysis of the $\kappa_4^{\crit}$ parameter measured on both sides of the transition, i.e., using data on both sides of the hysteresis region as close to the transition as one can get. This is presented in Fig.\ \ref{fig:k4_crit} where we compare fits of eq. (\ref{eq_scaling}) and eq. (\ref{eq_scaling_m1}) to these data. The vertical measurements are shown in the left panel of Fig. \ref{fig:k4_crit} and the corresponding  $\chi^2$ values in Table \ref{table:2}. The horizontal ones are in the  right panel of Fig. \ref{fig:k4_crit} and the $\chi^2$ values can be found in Table \ref{table:3}. Again, the difference between the two classes of fits is not large, however, our data favor slightly the first-order fits.

\begin{table}[h!]
\centering
\begin{tabular}{||c| c c| c c||}
 \hline
 Fit & $\chi^2_{\kappa_0 = 4.0,C}$ & $\chi^2_{\kappa_0 = 4.0,B}$ & $\chi^2_{\kappa_0 = 4.2,C}$ & $\chi^2_{\kappa_0 = 4.2,B}$ \\ [0.5ex] 
 \hline\hline
 free $\gamma$ & 1.15e-06 & 5.2e-05 & 1.3e-06 &  6.3e-06  \\
 $\gamma = 1$ & 1.05e-06 & 1.6e-05 & 8.1e-07 & 4.2e-06 \\ [1ex] 
 \hline
\end{tabular}
\caption{The $\chi^2$ values related to the individual fits of  eq. (\ref{eq_scaling}) (free $\gamma$) and  eq. (\ref{eq_scaling_m1}) ($\gamma = 1$) to finite volume scaling of the $\kappa_4^{\crit}$ in vertical measurements ($\kappa_0$ fixed). The $C$ or $B$ indices indicate  which side of the phase transition the fitted data belong to.}
\label{table:2}
\end{table}

\begin{table}[h!]
\centering
\begin{tabular}{||c| c c| c c||}
 \hline
 Fit & $\chi^2_{\Delta = 0,C}$ & $\chi^2_{\Delta = 0,B}$ & $\chi^2_{\Delta = -0.02,C}$ & $\chi^2_{\Delta = -0.02,B}$ \\ [0.5ex] 
 \hline\hline
 free $\gamma$ & 1.2e-04 & 2.2e-04 & 1.6e-04 &  4.1e-04  \\
 $\gamma = 1$ & 1.2e-04 & 1.0e-04 & 2.1e-04 & 3.8e-04 \\ [1ex] 
 \hline
\end{tabular}
\caption{The $\chi^2$ values related to the individual fits of  eq. (\ref{eq_scaling}) (free $\gamma$) and  eq. (\ref{eq_scaling_m1}) ($\gamma = 1$) to finite volume scaling of the $\kappa_4^{\crit}$ in horizontal measurements ($\Delta$ fixed). The $C$ or $B$ indices indicate  which side of the phase transition the fitted data belong to.}
\label{table:3}
\end{table}

Summarizing, we have shown that the first-order type of critical scaling relation (with the proposed finite size correction) fits our data at least as well or even better than the (standard) higher-order fits. Therefore, taking into account the behavior of order parameters which, as already shown in \cite{towardsUV} and discussed above, on both sides of the transition behave discontinuously even when extrapolated to the infinite volume limit, we conclude that the data favor a first-order $B\mi C$ transition, but with the admittedly strange  property that in the limit $\bar{N}_{41} \to \infty$ the region of hysteresis shrinks to zero.

\section{Discussion and conjecture}\label{discussion}

In this paper, we have analyzed the previously not measured $A\mi B$ phase transition at three different locations of fixed $\kappa_0 = 4.8$, $4.6$, and $4.5$. The finite size scaling analysis of $\Delta^{\crit}(\bar{N}_{41})$ revealed a scaling exponent $\gamma$ close to value $1$, which points to a first-order phase transition. Next, we analyzed the $B\mi C$ phase transition, improving the statistics and extending the previous research to another fixed value of the $\kappa_0 = 4.2$. We also analyzed the finite size scaling when approaching the critical line along the  perpendicular direction in the coupling constant space where  $\Delta$ was fixed at $0$ and $- 0.02$. We concluded that the $B\mi C$ phase transition was also a first-order transition, although we had to use a slightly modified fitting function to obtain acceptable agreement with data. Let us stress that the phase transitions considered are non-standard from the point of view of ordinary phase transitions, since they involve changes in spacetime, rather than changes in field configurations on a fixed spacetime. From the above and our earlier measurements, an interesting pattern emerges, relating the order of the CDT phase transition to changes in the topology of spacetime and it leads to the following conjecture: \textit{phase transitions which involve a change in topology will be first-order transitions}. 

First, let us clarify that by topology we mean \textit{effective} topology. This is exemplified by the configurations we observe in the $C$ phase if the topology of space is $S^3$ and we have time-periodic boundary conditions. Thus, from the outset the topology of the triangulations is that of $T^1 \times S^3$. This is encoded in the initial configuration when we start our MC simulations. By construction, any MC update of the configuration does not change the imposed topology. However, after many MC updates, a typical generic configuration will effectively look like an $S^4$-triangulation where the \textit{north-pole} and the \textit{south-pole} of this $S^4$ are connected by a thin \textit{stalk} of cut-off scale. While technically the triangulation still has the topology $T^1\times S^3$, it is clear that if allowed by the updating algorithm, the preferred configurations would have the topology of $S^4$. Secondly, recall from the discussion of the nature of generic configurations in Sec.\ \ref{sec-phasediagram}, that we observed a second-order transition between phase B and phase Cb, where there is no change in the effective topology. Likewise, in the case when the spatial topology is $S^3$ there is no real change in topology when we move from phase $C_b$ to phase $C$ and again we observe a second-order transition. All other phase transitions observed are first-order transitions and they are related to a change of the effective topology. In the case of toroidal spatial topology, the $C_b-C$ transition involves a change from $S^4$ in phase $C_b$ to $T^4$ in phase $C$. Similarly, the $B-C$ transition involves a change from $S^4$ to $T^4$. The effective topology of phase $A$ may be best characterized as the topology of a disjoint union of spatial geometries of various extensions (see upper left panel in Fig.\ \ref{fig:volprofs}). Thus, it is different from the topologies encountered in phase $C$ and phase $B$. Again, the observed phase transitions between phase $C$ and phase $A$, as well as the transition between phase $B$ and phase $A$, are first-order transitions.

With hindsight, it is not surprising that a change in the topology of spacetime at the phase transition might result in a first-order transition, since often one would need some major rearrangements of the configurations to implement the change. In this way, a barrier for such a change can be created and lead to pronounced hysteresis. Such phenomena are of course well known in field theories, where the topology of field configurations might change. The new thing here is that we are discussing the topology of spacetime itself. We were alerted to the relation between topology change of spacetime and phase transitions when we studied scalar fields coupled to geometry \cite{scalar2,scalar1}, but the phenomena seen there are in fact already present in the pure theory of geometry as discussed here.

In the CDT theory, our main interest is to find a second-order phase transition where we might be able to define a continuum limit and maybe even a UV fixed point. Since only phase $C$ seems to offer acceptable \textit{infrared} configurations, it is natural to look for phase transition lines in or at the border of phase $C$. From the results reported here, we cannot use toroidal spatial topology in such studies, since the transitions will be first-order. One should note that there is still a chance that even though the phase transition lines bordering the $C$ phase are first-order, the triple points where the lines end and meet may be of higher-order. This may be indeed the case of the $C_b\mi B\mi C$ triple point as the higher-order $B-C_b$ transition line ends in this point as well. Nevertheless, this scenario is not very natural and it is hard to be proven numerically. However, if the spatial topology  is $S^3$, the $C_b\mi C$ transition is a second-order transition. In addition it is then possible that the $B\mi C$ transition will also be second-order and it makes the $C_b\mi B\mi C$ triple point an interesting candidate for a UV-fixed point. At least it would then be a point where three second-order transition lines meet.  In addition, the $C_b-C$ transition has in that case the simple interpretation as the phase transition related to the breaking of isotropy and homogeneity. One could then imagine that the shadow of such a breaking could be important for inhomogeneities in our universe.

\section*{Acknowledgements}
J.G-S. and A.G. acknowledge support from the grant 2019/33/B/ST2/00589 from National Science Centre Poland. This research was funded by the Priority Research Area Digiworld under the program Excellence Initiative – Research University at the Jagiellonian University in Krakow. We would also like to acknowledge our deep gratitude to Prof. Jerzy Jurkiewicz (from Institute of Theoretical Physics of the Jagiellonian University in Krakow) who collaborated with us in this research; unfortunately he unexpectedly passed away on 30th November 2021, before we finished this work.

\begin{appendices}
\renewcommand{\theequation}{A-\arabic{equation}}
\setcounter{equation}{0}

\section*{Appendix 1: Machine learning  locations of the CDT phase transitions.}\label{sec:App1}

As already noted in Section \ref{sec:method}, in all phase transition studies described herein we have tested and used a new method of finding locations of the (pseudo-)critical points based on machine learning techniques. We have tried many {machine learning} (ML) methods\footnote{Detailed results of using Machine Learning in CDT data analysis will be published in a separate article. Here we just focus on the Logistic Regression used in the current study.} but the one which turned out to be both very simple and at the same time most efficient has been based on a {\it logistic regression} model. A detailed discussion of the logistic regression in the context of ML can be found, e.g., in \cite{machine}, so below we only shortly summarise this approach and describe most important aspects of its implementation in the context studied here.

Logistic regression  is a  so-called  supervised ML technique  which can be applied when one has labeled training data sets. It is commonly  used in  classification problems, where some new, yet unlabeled, data have to be divided into sub-classes attributed with labels of the training  set. A simple example studied here concerns data  measured in Monte Carlo simulations which can be classified  as belonging to one of two different phases, labeled  $P_0$ and $P_1$. Each data point $p_j$ within the MC measurements is characterized by a set of features (observables): $\{ x_i(p_j)\}$. In the binary classification problem, where one has only two classes, the probability that a data point $p_j$  belongs to class  $P_1$  is estimated by a logistic function: 
\begin{equation}\label{logprob}
\mathrm{Pr}(p_j\in P_1)=\frac{1}{1+b^{ -\sum_i w_i x_i(p_j)}}
\end{equation}
and the probability that $p_j$ belongs to class $P_0$ is then $\mathrm{Pr}(p_j\in P_0)=1 - \mathrm{Pr}(p_j\in P_1)$.
In the learning process a numerical algorithm optimises parameters of the logistic function, i.e.,  $b$ and  $w_i$ (weights), by minimizing a \textit{cost  function}, usually defined as the cross-entropy of the (labeled) training data set\footnote{Often, the learning data are additionally split into the \textit{training} and the \textit{test} sets. Parameter optimization (fitting) is done using the training set but model accuracy is checked using the test set.}:
\begin{equation}
\textrm{cost\ function} = -\frac{1}{N}\sum_{j=1}^N\big( l_j \ln \mathrm{Pr}(p_j\in P_1)+ (1-l_j) \ln \mathrm{Pr}(p_j\in P_0)\big),
\end{equation}
where $l_j=1$ if $p_j\in P_1$ and $l_j=0$ otherwise. In practice one  usually adds $L_1 =\lambda_1 \sum_i | w_i |$ (Lasso regularization) or $L_2 =\lambda_2 \sum_i w_i^2 $ (Ridge regularization) to the {\it cost function}, where $\lambda_1$ and $\lambda_2$ are metaparameters of the model. Including non-zero $\lambda_1$ or $\lambda_2$  reduces the number of non-zero weights $w_i$ which effectively  acts as a penalty for the number of parameters and thus enhances model prediction accuracy and interpretability.\footnote{The metaparameters $\lambda_1$ and $\lambda_2$ can themselves be optimized by fitting many models with different $\lambda_1$ and $\lambda_2$ and choosing the one with the highest accuracy. In all cases studied here the optimal models had very small values of these metaparameters.}  

\begin{figure}[h!]
\centering
\includegraphics[width=0.7\textwidth]{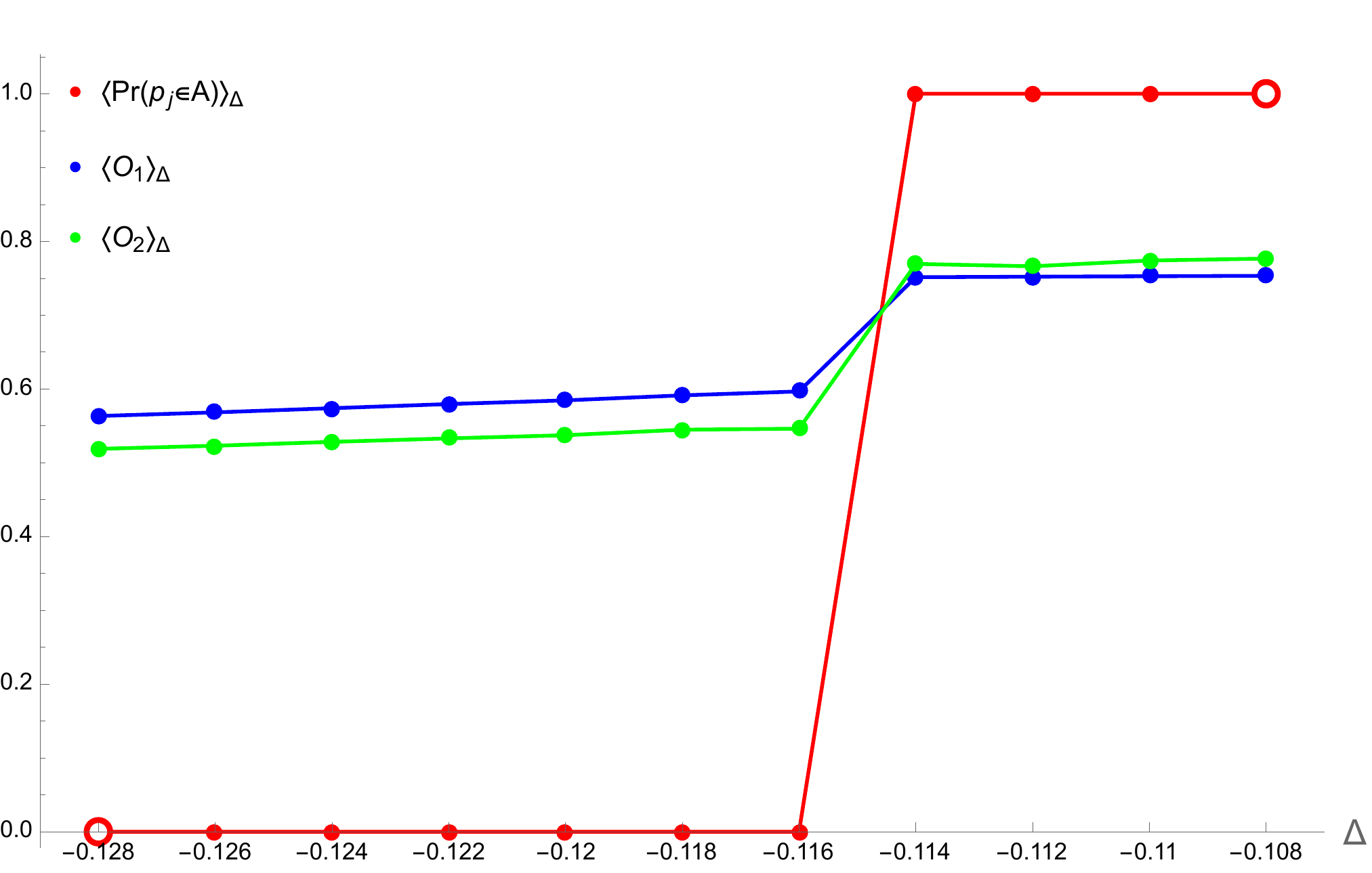}
\caption{Machine learning analysis of the $A\mi B$ transition for fixed $\kappa_0=4.8$ and $\bar N_{41}=100\mathrm{k}$.
Red points are the mean probabilities $\langle \mathrm{Pr}(p_j\in P_1)\rangle_\Delta$ that all  MC measurements performed for a given value of $\Delta$  belong to phase $A$. The probabilities were computed using the logistic regression model (\ref{logprob}) learned (fitted) using only part of data measured for the lowest and the highest $\Delta$ (denoted by empty dots). For comparison we also plot the \textit{traditional} order parameters $\OP_1$ (rescaled by $5\times$) and $\OP_2$ (rescaled by $35\times$).}
\label{fig_prob_AB}
\end{figure}

For each phase transition studied here, we  create a training data set by \textit{manually} attributing phase labels to CDT data measured deep into each phase and use these data to learn (fit) the logistic regression  model. Then we use the learned model to classify other data, measured closer to the transition region, as belonging to one of the two phases. The phase transition point is signalled by a sudden change of the probability (\ref{logprob}) of the measured data  to belong to a given phase from $\sim 0$ to $\sim 1$, see Fig.~\ref{fig_prob_AB}. We have decided to look at data measured for each chosen value of the fixed coupling constant and each lattice volume independently, i.e., for each such a choice we create an independent training set, fit our model and use it to find the critical point. Take, e.g., the $A\mi B$ transition analysis done for some fixed value of $\kappa_0$ and fixed $\bar{N}_{41}$ volume based on data measured for different values of the $\Delta$ coupling : $\Delta_{\mathrm{min}}, \Delta_1,....,\Delta_k, \Delta_{\mathrm{max}}$, see Section \ref{sec:AB}. The procedure to find $\Delta^{\crit}(\bar N_{41})$ is the following:\footnote{For the $B\mi C$ transition, see Section \ref{sec:BC},  $P_1$ denotes "phase C", and in the \textit{vertical} measurements the procedure is exactly the same while in the \textit{horizontal} measurements one fixes $\Delta$ and changes $\kappa_0$  (thus $\kappa_0 \leftrightarrow \Delta$).}
\begin{enumerate}
\item take all data measured for the highest value of $\Delta=\Delta_{\mathrm{max}}$ (deep in phase $A$) and label it as $P_1$ (or "phase $A$"),
\item take all data measured for the lowest value of $\Delta=\Delta_{\mathrm{min}}$ (deep in phase $B$) and label it as $P_0$ (or "phase $B$"),
\item take a subset of data from points 1 and 2 as a (labeled) {\it  training set},
\item learn (fit) the logistic regression model using the training set\footnote{During ML process we have chosen not to split the learning data into independent  training and test sets. Instead we make a global check of the model accuracy - see point 5.}; this step also includes optimizing metaparameters of the model ($\lambda_1$ and $\lambda_2$ regularization coefficients),
\item check  accuracy of the fitted model using all data from points 1 and 2 (in all cases analyzed the accuracy was 100\%, i.e., the model was able to distinguish between the $P_1$ and $P_2$ data perfectly),
\item use the fitted model to  classify  other, yet unlabeled, data measured for $\Delta$:  $\Delta_1,....,\Delta_k$, i.e., closer to a phase transition region than data from points 1 and 2,
\item compute the probability of all classified data points to be in phase $P_1$:  $\mathrm{Pr}(p_j\in P_1)$, see eq. (\ref{logprob}),
\item compute the mean value $\langle \mathrm{Pr}(p_j\in P_1)\rangle_\Delta$ (and the error: standard deviation) of the probability of all data points measured for a given $\Delta$, 
\item find the transition point $\Delta^{\crit}(\bar{N}_{41})$,  where $\langle \mathrm{Pr}(p_j\in P_1)\rangle_\Delta$ jumps  from $\sim 0$ to $\sim 1$, see Fig.~\ref{fig_prob_AB}.
\end{enumerate} 
 Then we repeat the analysis (model learning and classification) for a different lattice volume $\bar N_{41}$, and then also for each different $\kappa_0$ independently.
 
 It is important to stress that in the data analysis described above we used only geometric information about triangulations (configurations)  generated in MC simulations, i.e., the set of  features used in the ML model fitting and classification  does not include any information about the values of the CDT coupling constants or other MC simulation parameters. Therefore we can say, that the method learned to distinguish the phases purely on (some) geometric features of the triangulations. We used a set of $30$ geometric observables, including both global parameters of the triangulations: $N_0$, $N_1$, $N_2$, $N_4$, $N_{41}$, $MO$, i.e., the total number of vertices, links, triangles, four-simplices, $\{4, 1\}$-simplices and maximal coordination number of vertices, respectively, as well as "per-time-slab" data: $N_{41}(t)$, $N_{32}(t)$, $N_{23}(t)$, $N_{14}(t)$, $N_0(t)$, $MO(t)$, i.e., the number of $\{4, 1\}$, $\{3, 2\}$, $\{2, 3\}$ and $\{1, 4\}$-simplices in a given slab between (lattice) time coordinate $t$ and $t+1$, the number and maximal coordination number of all vertices with a time coordinate $t$, respectively. In all cases $t = 1,2,3,4$ (with periodic boundary conditions). In order to increase statistics of our data set and also to encode information about a time shift symmetry of the CDT model, we have quadrupled the data by performing a time shift of all "per-time-slab" features by (periodically)  changing their time  coordinates $t = (1,2,3,4) \to (4,1,2,3) \to (3,4,1,2) \to  (2,3,4,1) $. The values of the global features kept unchanged.

The above-mentioned observables also enable one to compute \textit{traditional} order parameters, which were earlier used to distinguish between the CDT phases, e.g., $\OP_1\equiv N_0/N_{41}$ or $\OP_2\equiv N_{32}/N_{41}$ and thus  check if the automatic ML classification algorithm is consistent with our former results, see Fig. \ref{fig_prob_AB}. We implemented the logistic regression model in \textit{Wolfram Mathematica 12} using the built-in function:
\begin{center}
\verb
Classify[..., Method -> "LogisticRegression"]
\end{center}  
with standard parameters. In all cases analyzed the model classified both $A\mi B$ and $C \mi B$ phase transition data correctly, with a $100\%$ accuracy, see Fig.~\ref{fig_prob_AB} as an example.

\end{appendices}

\end{document}